\documentclass[a4paper，]{cas-dc}
\def\tsc#1{\csdef{#1}{\textsc{\lowercase{#1}}\xspace}}
\tsc{WGM}
\tsc{QE}

\usepackage{dblfnote}
\usepackage{svg} 

\usepackage{amsmath,amssymb,amsfonts,amsthm}
\usepackage{algorithmic}
\usepackage{graphicx}
\usepackage{array}
\usepackage{ulem}
\usepackage{textcomp}
\usepackage{xcolor}
\usepackage[T1]{fontenc}
\usepackage{makecell}
\newtheorem{definition}{Definition}
\usepackage{booktabs}
\usepackage{multicol,multirow}
\usepackage{stfloats}
\usepackage[numbers,sort&compress]{natbib}

\begin{document}
\let\WriteBookmarks\relax
\def\floatpagepagefraction{1}
\def\textpagefraction{.001}
\let\printorcid\relax 

\shorttitle{}    

\shortauthors{C. An et al.}

\title[mode = title]{Incorporating Like-Minded Peers to Overcome Friend Data Sparsity in Session-Based Social Recommendations}  

\author[1]{Chunyan An}
\author[1]{Yunhan Li}
\author[2]{Qiang Yang}\cormark[1]
\author[3]{Winston K.G. Seah}
\author[4]{Zhixu Li}
\author[1]{Conghao Yang}

\address[1]{College of Computer Science, Inner Mongolia University, Hohhot, Inner Mongolia, China} 
\address[2]{College of Medicine, University of Florida, Gainesville, USA} 
\address[3]{School of Engineering and Computer Science, Victoria University of Wellington, Wellington, New Zealand} 
\address[4]{School of Computer Science, Fudan University, Shanghai, China} 
\cortext[1]{Corresponding author at: College of Medicine, University of Florida, Gainesville, USA.
\\
E-mail address: ann@imu.edu.cn (C. An), 32109154@mail.imu.edu.cn
(Y. Li), qiangyang@ufl.edu (Q. Yang).}  

\begin{abstract}
Session-based Social Recommendation (SSR) leverages social relationships within online networks to enhance the performance of Session-based Recommendation (SR). However, existing SSR algorithms often encounter the challenge of ``friend data sparsity''. Moreover, significant discrepancies can exist between the purchase preferences of social network friends and those of the target user, reducing the influence of friends relative to the target user's own preferences.
To address these challenges, this paper introduces the concept of ``Like-minded Peers'' (LMP), representing users whose preferences align with the target user's current session based on their historical sessions. This is the first work, to our knowledge, that uses LMP to enhance the modeling of social influence in SSR. This approach not only alleviates the problem of friend data sparsity but also effectively incorporates users with similar preferences to the target user.
We propose a novel model named Transformer Encoder with Graph Attention Aggregator Recommendation (TEGAARec), which includes the TEGAA module and the GAT-based social aggregation module. The TEGAA module captures and merges both long-term and short-term interests for target users and LMP users. Concurrently, the GAT-based social aggregation module is designed to aggregate the target users' dynamic interests and social influence in a weighted manner. 
Extensive experiments on four real-world datasets demonstrate the efficacy and superiority of our proposed model and ablation studies are done to illustrate the contributions of each component in TEGAARec.
\end{abstract}




\maketitle

\section{Introduction}

Session-based recommendation (SR) is tasked with modeling a user's interest preferences based on historical data, such as the user's past purchases within a given timeframe, and predicting the next item of interest \cite{wang2021survey,zhao2023Attn,wu2019session,wang2022cgsnet,liu2023semantic}. Previous studies have demonstrated that incorporating users' social relationships can enhance recommendation performance \cite{feng2021hierarchical,tang2022time, zhao2023social1, long2021social2}. When social relationships are integrated into the session-based recommendation task, it becomes known as session-based social recommendation (SSR) \cite{chen2021efficient,xiao2017learning}.

Numerous methods have been proposed to enhance the performance of SSR tasks, which can be broadly categorized into two groups, i.e.,  historical behavioral information based methods \cite{yuan2019simple, da2021adaptive, du2023sequential, liu2018stamp, wu2019session, wang2020global, zhang2023efficiently} and social/interactive information based methods \cite{chen2021efficient, song2019session,li2017towards, liu2023gnnrec}. 
The first group focuses on leveraging historical behavioral information to predict the recommendation information of the target user. Early methods included collaborative filtering techniques \cite{linden2003amazon}, while more recent approaches focus on treating session data as sequences and modeling them using recurrent neural networks (RNNs) \cite{linden2003amazon,hidasi2015session,li2017neural,medsker2001recurrent}. 
The second group centers on information aggregation, wherein the target user's embedded representation is learned by integrating information from both the user-user social network and the user-item interaction network \cite{song2019session,chen2021efficient,liu2023gnnrec}. Many existing methods adopt Graph Neural Networks (GNNs) to construct their models, aiming to capture the intricate item transitions within sessions \cite{scarselli2008GNN,vaswani2017attention, wan2023spatio,velivckovic2017graph,medsker2001recurrent}.

However, real-world SSR algorithms encounter several challenges. Firstly, the issue of ``friend data sparsity'' arises, where the recommendation data may lack sufficient information about the friends of the target user, resulting in lost interaction data in the current or historical sessions. Secondly, the preferences of social network friends may differ significantly from those of the target user, and the influence of friends may be limited compared to the target user's own preferences especially facing the problem of friend data sparsity, potentially leading to inaccurate recommendation. Last but not the least, users' interests may evolve dynamically over time, posing a considerable challenge in accurately modeling interest changes of users \cite{wang2021survey,song2019session,velivckovic2017graph}.

\begin{figure}
	\centering
        \includegraphics[width=1.0\linewidth, keepaspectratio]{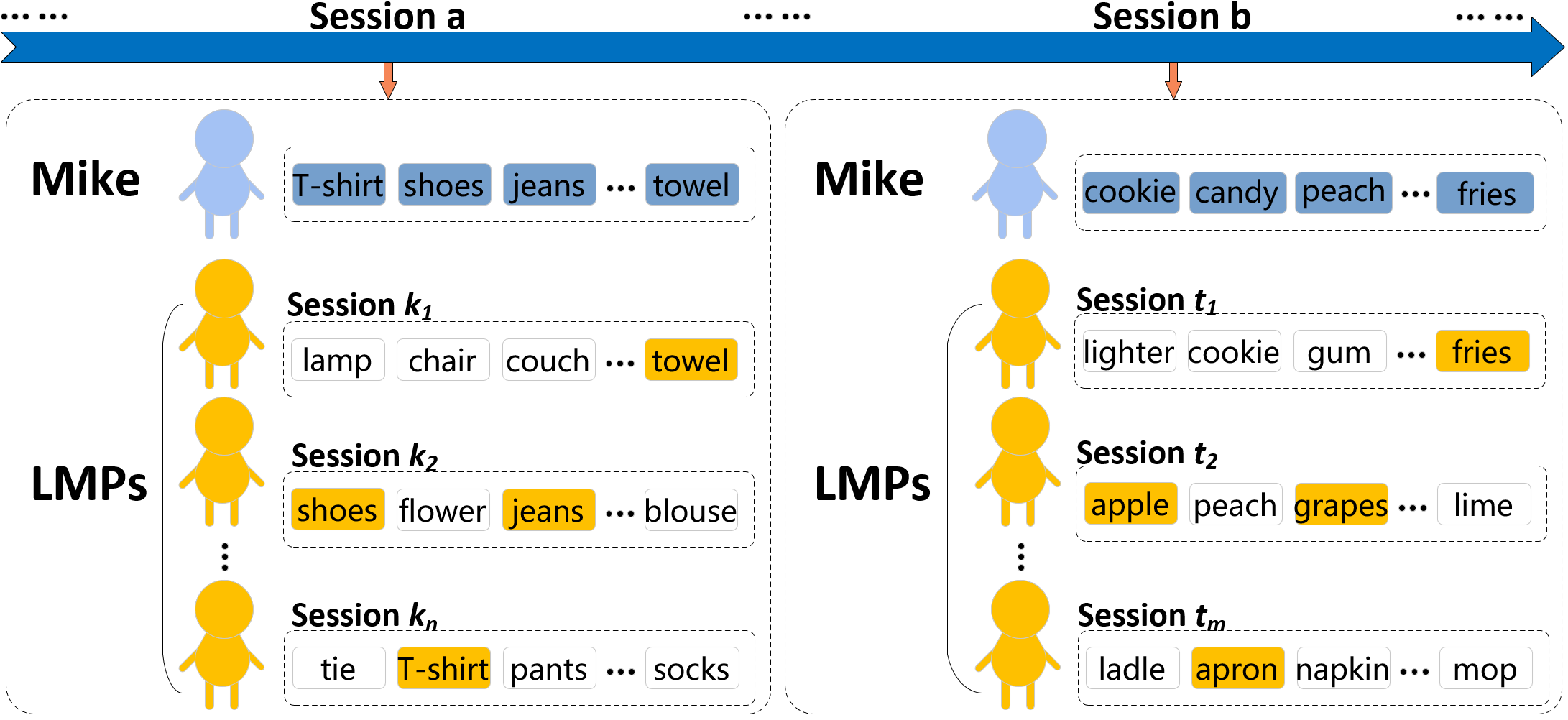}
	\caption{Like-minded Peers (LMP) mined by Mike in different sessions -- Session $k_x$ and Session $t_x$ denote that the LMP has interacted with the same items as Mike in Session $a$ and Session $b$ respectively, where $k_x \in [1, a-1 ]$ and $t_x \in [1, b-1]$.}
	\label{fig:1}
\end{figure}

To tackle these challenges, we propose the concept of ``Like-minded Peers'' (LMP) to represent users whose preferences are aligned with the target user's current session in their history sessions. We posit that users who exhibit similarity in interaction patterns with the target user, even if they are not social friends, can offer valuable insights. To the best of our knowledge, this is the first work using LMP to enhance SSR. It not only alleviates the problem of friend data sparseness but also effectively incorporates users with similar preferences to the target user.
For example, consider a scenario where a mother purchases diapers and formula. In this case, she is more likely to share interests with other users who have also bought diapers and formula in the past. This alignment is attributed to the fact that social friends may not necessarily share the same motherhood identity, thus their purchasing behavior has a lesser influence on her buying propensity.
To promote the understanding of LMP, we also provide an 
example, as shown in Fig. \ref{fig:1}. Given the target user Mike in different sessions, at the moment of session a/b, his LMP is defined as the other users who have interacted with Mike for the same item in session a/b before that.
It is notable that while the concept of LMP bears similarities to user-based collaborative filtering algorithms, traditional collaborative filtering methods often suffer from data sparsity issues and limited model expressiveness \cite{sharma2013survey}. 
In this paper, we collectively refer to the target user's LMP users and social friends as the target user's $neighbours$. 

In addition, considering that the interests of the users (including LMP users and social friends) are dynamic over time, mainstream approaches typically focus only on short-term interests in the current session \cite{yuan2019simple, li2017neural, liu2018stamp, wu2019session, wang2020global,li2022enhancing, zhang2023efficiently}, or consider only the long- and short-term dynamic interests of neighbours, ignoring the long-term interests of the target user \cite{song2019session, liu2023gnnrec}. In this paper, we model the long-term and short-term interests of the target users as well as their neighbours (including LMP users and social friends) to better capture their dynamic interests. Particularly, we propose a novel model called \textbf{T}ransformer \textbf{E}ncoder with \textbf{G}raph \textbf{A}ttention \textbf{A}ggregator \textbf{Rec}ommendation (TEGAARec), consisting of a TEGAA module and a graph attention (GAT)-based social aggregator module. TEGAA module focuses on modeling and fusing the dynamic interests of the target user and her/his neighbours. It contains a Transformer \cite{vaswani2017attention} encoder to encode the short-term interests, a long-term interests encoder to get the long-term interests and a GAT-based Dynamic Interest Aggregator to aggregate the information of items. The learned representations are fed into a GAT-based Social Aggregation module to weightedly aggregate the target users' dynamic interests and the social influence.

The main contributions of this paper are as follows:
\begin{itemize}
    \item[$\bullet$] To the best of our knowledge, we are the first work to introduce the concept of ``Like-minded Peers'' (LMP) to alleviate the influence of ``friend data sparsity'' in SSR .
    \item[$\bullet$] We propose a novel TEGAARec model, which can model the dynamic interests of the target user and her/his neighbours separately through our designed TEGAA module. A GAT-based Social aggregation module is proposed to weightedly aggregate the target users' dynamic interests and the social influence.
    \item[$\bullet$] We conduct extensive experiments on four real-world datasets, which shows a large improvement compared to the state-of-the-art work at several metrics. 
\end{itemize}

\section{Related Work}

\subsection{Session-based Recommendation}
Session-based recommendation tasks involve sequential recommendation, where early approaches predominantly utilized models from the Recurrent Neural Network (RNN) family to capture temporal order \cite{wang2021survey}. For example, Hidasi et al. \cite{hidasi2015session} proposed to use a variant of RNN, namely the Gated Recurrent Unit (GRU) \cite{chung2014gru}, to treat sessions as sequence prediction tasks for session-based recommendation. Li et al. \cite{li2017neural} further improved upon this by incorporating attention mechanisms into the GRU to better capture users' sequential behaviors and personal intentions.

GNN-based methods have been introduced to address the limitations of RNN models in capturing complex item transitions within sessions. Wu et al. \cite{wu2019session} proposed the SR-GNN model, which represents sessions as directed graphs and learns item transitions using Gated Graph Neural Networks (GGNN) \cite{li2015GGNN}. Xu et al. \cite{xu2019graph} proposed a GC-SAN model that introduces an attention mechanism to learn global dependencies on top of SR-GNN to enhance the session representation. Li et al. \cite{li2022enhancing} proposed a HIDE model that goes a step further by introducing a hypergraph neural network approach to the SR task, which models possible interest transitions from different perspectives, and then uses a combination of micro- and macro-approaches to learn the intent behind each clicked item.
However, the increasing complexity of GNN-based approaches has led to only marginal improvements. Zhang et al. \cite{zhang2023efficiently} introduced the Attn-mixer, a multilevel attention mixer network that achieves multilevel reasoning for item transitions without relying on GNNs. Wang et al. \cite{wang2020global} proposed the GCE-GNN model, which significantly enhances recommendation performance by constructing a global graph to learn item-transition information from other sessions. 

Despite their successes, these approaches often overlook the modeling of users' dynamic interests and lacks the modeling of long-term interests across sessions. In contrast, our approach integrates both long-term and short-term user interests to achieve a dynamic representation of users' interests.

\subsection{Social Recommendation}
User-based collaborative filtering (CF) models \cite{linden2003amazon, luo2008collaborative, jia2015user} are the classical algorithm to recommend items to the target user based on user similarity by finding other users with similar preferences or purchase history to the target user and recommending items to the target user that have been liked or purchased by these similar users. They operate under the assumption that users who have interacted with similar items in the past tend to share similar preferences, making information about such users valuable for recommendation. However, they often encounter problems of data sparsity and limited model expressiveness \cite{wang2021survey,sharma2013survey}.

\subsection{Session-based Social Recommendation}
Several existing methods tried to address friend data sparsity problem by employing repeated friend selection \cite{song2019session,liu2023gnnrec}.
For instance, Song et al. \cite{song2019session} proposed the DGRec model, which utilized RNN-based models to capture the dynamic interests of users and their friends, while employing graph attention networks to model the influence of friends. Building upon DGRec, Liu et al. \cite{liu2023gnnrec} introduced GNNRec, an enhanced model that constructed session graphs and leveraged Gated Graph Neural Networks (GGNN) to model complex item transitions, further improving recommendation performance. 
Additionally, Chen et al. \cite{chen2021efficient} proposed SERec, a generalized social recommendation framework that utilized heterogeneous graph neural networks to learn user-item representations, effectively capturing item transitions across sessions. Feng et al. \cite{feng2021hierarchical} proposed a Hierarchical Social Similarity-guided Model with Dual-mode Attention (HMDA) for Session-based Recommendation where the social influence exerted by friends are aggregated.

Although these methods perform well, they are less effective when the user has fewer friends. To overcome this limitation, we introduce the concept of ``Like-minded Peers'' (LMP), which can enable the identification of users who genuinely share similar preferences with the target users.

\section{Problem Definition}

Before delving into the problem definition, we first give the basic definitions of concepts used throughout this paper and then the definitions relevant to users: Like-minded Peers (LMP), social friends, and neighbours. 

We denote $G=(U,E)$ as the social network where $U$ and $E$ are the users and the social relations among users, respectively. $e_{(\bar{u},u)}=1$ represents the existence of a social edge between the target user $\bar{u}$ and user ${u}$, otherwise $e_{(\bar{u},u)}=0$. $Q$ represents the set of all sessions of users, and ${Q}_u = \{S_{1}^{u}, S_{2}^{u},\dots, S_T^u \}$ denotes the set of sessions for a user $u \in U$, where $S_{t}^{u}$ is the $t$-th session of user $u$. We denoted $S_{t}^{u} = \{i_{t,1}^u, i_{t,2}^u, \dots, i_{t,n}^u\}$ as the set of $n$ items that the user $u$ preferred at the $t$-th session. $T$ is the number of historical sessions of user, and $n$ is the number of items in the session. Each user's sessions are chronologically arranged.

\begin{definition}
\textbf{Like-Minded Peers.} Given a target user $\bar{u}$ in the $T+1$-th session, the Like-Minded Peers ($\mathcal{X}_{{\bar{u}}}^{T+1}$) of $\bar{u}$ are the set of users ranging from the historical session interval $[1,T]$ who have item intersections with the user $\bar{u}$ in session $T+1$. It is formally defined below:
\begin{equation}
    \begin{aligned}
        \mathcal{X}_{{\bar{u}}}^{T+1} &=\{\forall u, u \in U\}\\  
        &\mbox{s.t.} \quad S_{T+1}^{\bar{u}} \cap \bigcup_{t=1}^{T}S_{t}^{u} \ne \varnothing, &\\
       & u \neq \bar{u} &\\
    \end{aligned}
    \label{eq3}
    \nonumber
\end{equation}
\end{definition}

\begin{definition}
  \textbf{Social Friends.} Given a target user $\bar{u}$ in the $T+1$-th session, social friends ($\mathcal{Z}_{{\bar{u}}}^{T+1}$) are the set of users ranging from the historical session interval $[1,T]$ who have connection edge in the social network $G$ with the target user $\bar{u}$ and have interaction behaviors in the historical session. It is formally defined below:
\begin{equation}
    \begin{aligned}
        \mathcal{Z}_{{\bar{u}}}^{T+1} &=\{\forall u, u \in U\}\\  
        &\mbox{s.t.} \quad e_{(\bar{u},u)}=1, \bigcup_{t=1}^{T}S_{t}^{u} \ne \varnothing, &\\
       & u \neq \bar{u} &\\
    \end{aligned}
    \label{eq3}
    \nonumber
\end{equation}
\end{definition}

\begin{definition}
   \textbf{Neighbours of Users.} Traditionally, neighbours are the users who directly/indirectly connect the target user in the social network. In this paper, we give a broad definition of neighbour which is the union set of LMP users $\mathcal{X}_{{\bar{u}}}^{T+1}$ and social friends $\mathcal{Z}_{{\bar{u}}}^{T+1}$ for the target user $\bar{u}$, denoted as:
   \begin{equation}
       \mathcal{N}_{{\bar{u}}}^{T+1} = \mathcal{X}_{{\bar{u}}}^{T+1} \cup \mathcal{Z}_{{\bar{u}}}^{T+1} 
       \nonumber
   \end{equation}
\end{definition}

\begin{definition}
    \textbf{Session-based Social Recommendation (SSR).} Given a new session $S_{T+1}^{\bar{u}}$ for target user $\bar{u}$ and her/his neighbours $\mathcal{N}_{{\bar{u}}}^{T+1}$, the goal of SSR is to recommend a set of items from $I$ that $\bar{u}$ is likely to be interested in during next time step $T+1$. Here, we employ the session information coming from the dynamic interests including the historical sessions $\bigcup_{t=1}^{T}S_{t}^{\bar{u}}$ and the current session $S_{T+1}^{\bar{u}}$ as well as the neighbour information including LMP users, i.e., $\mathcal{X}_{{\bar{u}}}^{T+1}$ and social friends i.e., $\mathcal{Z}_{{\bar{u}}}^{T+1}$. 
\end{definition}

\section{Proposed Method}
\begin{figure*}
	\centering
	\includegraphics[width=1.0\textwidth]{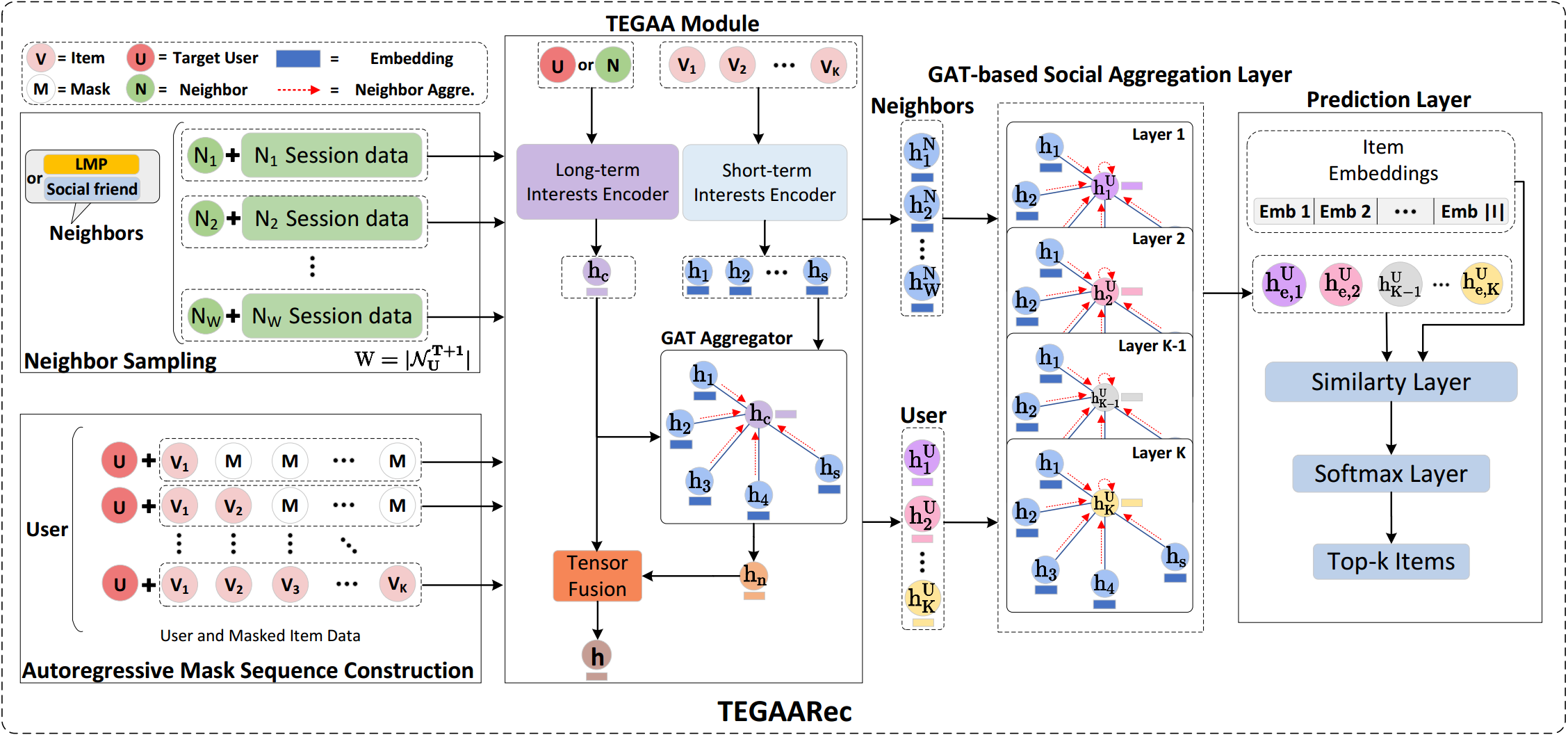}
	\caption{The overall framework of TEGAARec.}
	\label{fig:2}
\end{figure*}

In this section, we present the detail of our proposed TEGAARec model. Fig.~\ref{fig:2} illustrates the overall framework, which first generates neighbours by randomly sampling LMP users and social friends from the historical sessions, then encodes and merges the long- and short-term interests of the target user and neighbours with our proposed TEGAA module. Next, a GAT-based social aggregation layer is designed to weightedly aggregate the target users' dynamic interests and the social influence. Finally, a prediction layer is used to calculate the similarities between the item embeddings and target user embedding for selecting the top-$k$ items as the recommendation.

\subsection{Historical Session-based Neighbour Sampling}
In this section, we introduce how to get the neighbours  $\mathcal{N}_{{\bar{u}}}^{T+1}$, i.e., LMP users and social friends, of the target user $\bar{u}$ given the historical sessions $\bigcup_{t=1}^{T} S_{t}^{\bar{u}}$. Particularly, for the LMP users acquisition, we first collect the items of $\bar{u}$ at the current session $S_{T+1}^{\bar{u}}$ and then check the users from the historical sessions ranging from the historical session interval [1, T] where users sharing the same items with $\bar{u}$ are selected as the candidates. Then, we randomly choose users from candidates as the LMP users according to a predefined value called $L_l$. 
Note that if the size of LMP users is less than $L_l$, we randomly select duplicate LMP from the existing LMP set. Finally, the LMP users, i.e., $\mathcal{X}_{{\bar{u}}}^{T+1}$ are captured from the historical sessions. Note that our sampling strategy can be extended to the more complex one by replacing the random selection with the volume-based selection of shared items.
Similarly, for social friends, we first extract users from the social networks in the historical sessions. The users directly connected to $\bar{u}$ and having interaction behaviors in the historical session are treated as the candidates. Next, the random sampling strategy is employed to select users based on the predefined value $L_s$ 
to get $\mathcal{Z}_{{\bar{u}}}^{T+1}$ as the social users.
We union the LMP users and social friends to acquire neighbours of the target user $\bar{u}$.

\subsection{Autoregressive Mask Sequence Construction}
\label{sec:Encoding}
To augment the number of training examples \cite{song2019session, wu2019session}, we try to split the current session sequence $S_{T+1}^{\bar{u}}$ into several sub-sessions sequence $MaskedInput_{k}$ and the corresponding labels $Target_{k}$. Specifically, for the sequence of interaction item sequence $S_{T+1}^{\bar{u}} = \{ i_{T+1,1}^{\bar{u}},i_{T+1,2}^{\bar{u}},\dots,i_{T+1,n}^{\bar{u}} \}$, we utilize
an Autoregressive Mask Sequence Construction module to split it as follows:
\begin{equation}
	\begin{aligned}
		MaskedInput_{k} &= \{i_{T+1,1}^{\bar{u}}, \dots, i_{T+1,k}^{\bar{u}}, 0, \dots, 0\},\\
		Target_{k} &= \{i_{T+1,k+1}^{\bar{u}}\}
	\end{aligned}
\end{equation}
where $MaskedInput_k$ represents the $k$-th constructed masked input sequence and $Target_k$ represents the $k$-th item to be predicted, $k \in [1, n-1]$.
Different from existing methods which augment data in the data preprocessing step \cite{wu2019session, wang2020global}, we implement it during the model's training phase. Our strategy can avoid the repeated sampling of neighbours in the process of historical session-based neighbour sampling.

\subsection{Long- and Short-Term Interests Encoding and Fusing with TEGAA Module} \label{section c}
Generally, interests of users may vary over time, therefore, it is essential to encode the long- and short-term interests, which can systemically learn the preference of target users. Given any user $u_m$ either from the neighbours $\mathcal{N}_{{\bar{u}}}^{T+1}$ or from the target user $\bar{u}$, and her/his interacted item sequence $S_t^{u_m}  = \{i_{t,1}^{u_m}, i_{t,2}^{u_m}, ...,i_{t,n}^{u_m}\}$, we aim to learn the fused representations of long-term and short-term interests of the user $u_m$. 

\subsubsection{TEGAA Module}
To capture the long-term and short-term interests and merge these interests of an user $u_m$, we propose a novel TEGAA module which consists of a Transformer Encoder, a user-embedding look-up table and a GAT-based Dynamic Interest Aggregator. 

\textbf{Long- and Short-Term Interests Encoding.} Given the item interaction sequence $S_t^{u_m} =\{i_{t,1}^{u_m},i_{t,2}^{u_m},...,i_{t,n}^{u_m}\}$, we first use the Transformer Encoder to get the embeddings of $S_t^{u_m}$ as the short-term interests below:
\begin{equation}
    h_t^{u_m} = Transformer\_Encoder(S_t^{u_m})
\end{equation}
where $Transformer\_Encoder(\cdot)$ is the function to get the encoded information of the interaction sequence of the session $S_t^{u_m}$.
We employ a single vector from the user-embedded lookup table to learn the representation of the long-term interests of the user $u_m$ below:
\begin{equation}
	emb_{u_m} = Embedding_{user} (u_m)
\end{equation}
where $Embedding_{user} \in\mathbb{R}^{|U| \times d}$ denotes the user embedding lookup table, $|U|$ is the number of all users, and $d$ is the size of the embedding dimension.

\textbf{Long- and Short-Term Interests Fusing.} Given the encoded information of items and users, we first modify the Multi-head Attention Mechanism to obtain the Multi-head Graph Attention Mechanism (MHGAT) by adjusting the query vectors to the center node, i.e., the target user embedding, and the key vectors to the neighbour embedding. MHGAT is utilzied to capture the short-term interest of the user ${u_m}$.
%
Particularly, we treat the embeddings of target user as the center node and the embeddings of neighbours as the neighbouring nodes of multi-head attention. They are calculated as follows:
\begin{equation}
	Q= emb_{u_m}W_{i}^{Q},K=h_{t}^{u_m}W_{i}^{K},V=h_{t}^{u_m}
\end{equation}
\begin{equation}
	head_i = Attention(Q, K, V)= SoftMax(\frac{QK^{\intercal}}{\sqrt{d_k}})V\label{eq:attention}
\end{equation}
\begin{equation}
	\begin{aligned}
		h_s &= M\!H\!G\!AT(Q, K, V)\\&= Concat(head_1; \dots; head_z) W^O + b^O
  \label{eq:head}
	\end{aligned}
\end{equation}
where $
head_1,\dots, head_z
$ denotes the $z$ heads of the multi-head attention mechanism, $Concat$ denotes the vector splicing operation, 
${W}^{O}\in\ \mathbb{R}^{{z}{d}_{k}\times{d}}, {W}_{i}^{Q}\in\ \mathbb{R}^{{d}\times{d}_{k}}, {W}_{i}^{K}\in\ \mathbb{R}^{{d}\times{d}_{k}}$ are the mapping parameter matrices, $\ {b}^{O}\in\ \mathbb{R}^{d}$ is the bias term, $\intercal$ denotes the transpose operation, and ${d}_{k}=d/h$.

Then, we fuse the short-term interest result $h_s$ obtained from the Dynamic Interest Aggregator with the long-term interest representation $emb_{u_m}$ to obtain the final representation $h_{u_m}$ below:
\begin{equation}
        \begin{aligned}
            h_{u_m} &= TensorFusion(h_s,emb_{u_m})\\
                &= ReLU(Concat[h_s; emb_{u_m}] {W^F} + {b^F})
        \end{aligned}
\end{equation}
\noindent 
where $ReLU(x) = max(0, x)$ is the nonlinear activation function, ${W^F} \in \mathbb{R}^{2d\times d}$ is the mapping parameter matrix, ${b^F} \in \mathbb{R}^d$ is the bias term, and $d$ is the size of the embedding dimension.

\subsubsection{Neighbour and Target User Interests Encoding}
Given the designed TEGAA module, we apply it for the neighbour interests encoding and the target user interesting encoding to get the fused long-term and short-term interests. Particularly, for the neighbour interest encoding, we use the TEGAA module to get the encoded results for all $|\mathcal{N}_{{\bar{u}}}^{T+1}|$ of the neighbours in the $T$-th session $H_{n_m} = \{h_{n_m,1}, h_{n_m,2}, \dots, h_{n_m,|\mathbf{\mathcal{N}_{{\bar{u}}}^{T+1}}|}\}$
\begin{equation}
	h_{n_m,k}=TEGAA\_Module(S_T^k) 
\end{equation}
where $S_T^k$ represents the sequence of interaction items for the $k$-th neighbour's $T$-th session, $h_{n_m, k}$ represents the coded representation of neighbour $k$ in the neighbour set through the TEGAA module, 
and $k \in [1, |\mathcal{N}_{{\bar{u}}}^{T+1}|]$.

Similarly, for the target user interests encoding, we feed the constructed masked input sequences into the TEGAA module, and it outputs the encoded information corresponding to each masked sequence of the target user $\bar{u}$. 
The encoded embeddings of $n-1$ sub-sessions of the target user
$\bar{u}$ is denoted as $H_{\bar{u}}=\{h_{{\bar{u}},1}, h_{{\bar{u}},2}, \dots, h_{{\bar{u}},n-1}\}$. Each element $h_{{\bar{u}},k} \in H_{\bar{u}}$ is calculated below:
\begin{equation}
	h_{{\bar{u}},k} = TEGAA\_Module(MaskedInput_k)
\end{equation}
where $k$ is the length of the session.

\subsection{GAT-based Social Aggregation Layer}
\label{sec:GATL}
Due to the difference of social influence for neighbours for  target user, it is necessary to discern their contributions.  In this section, we consider to use the MHGAT module (as mentioned in section \ref{section c}) to weightedly aggregate the target users' dynamic interests and the social influence. On one hand, the neighbours of the target user $\bar{u}$ in are encoded in session $T$ with $H_{n_m} = \{h_{n_m,1}, h_{n_m,2}, \dots, h_{n_m,|\mathbf{\mathcal{N}_{{\bar{u}}}^{T+1}}|}\}$ as neighbour nodes. On the other hand, the encoding result $H_{\bar{u}}=\{h_{{\bar{u}},1}, h_{{\bar{u}},2}, \dots, h_{{\bar{u}},n-1}\}$ of each mask sequence in Session $T+1$ for the target user $\bar{u}$ is used as the target aggregation node. Node types of the target aggregation node and the neighbour nodes are both users, i.e.,  homogeneous nodes, so we add the self-attention mechanism to the aggregation. The final aggregated result is obtained as $H_{\bar{u}}^{\prime}= \{h_{{\bar{u}},1}^{\prime}, h_{{\bar{u}},2}^{\prime}, \dots, h_{{\bar{u}},n-1}^{\prime}\}$, whose element is calculated below:

\begin{equation}
h_{{\bar{u}},k}^{\prime}=M\!H\!G\!AT(h_{{\bar{u}}, k}W_{i}^{Q},h_{n_m^{\prime}}W_{i}^{K},h_{n_m^{\prime}})
\end{equation}
where $h_{n_m^{\prime}}=
[h_{n_m,1}, \dots, h_{n_m,|\mathcal{N}_{{\bar{u}}}^{T+1}\}|}, h_{{\bar{u}},k}]$
, and
$h_{{\bar{u}},k}$ represents the encoding result of the target user ${\bar{u}}$ for the $k$-th mask sequence in session $T+1$, $k \in  [1, n-1]$. 

\subsection{Prediction Layer}
In this section, we introduce how to select the top relevant items that the target user preferred.
We first encode all the items using an item-embedded lookup table. Given any item $i$ with the id $i_v$, we get its embedding below:
\begin{equation}
	h_v^i=Embedding_{item} (i_v)
\end{equation}
Then, we use the SoftMax function to predict the probability of the next item as follows:
\begin{equation}
	\begin{aligned}
		p(Target_k\mid &MaskedInput_k;\{S_t^{n_m},n_m\in \mathcal{N}_{{\bar{u}}}^{T+1}\})\\
		&=\frac{exp((h_{{\bar{u}},k}')^\intercal h_{Target_k}^i)}{\sum_{j=1}^{|I|}exp((h_{{\bar{u}},k}')^\intercal h_j^i)}
	\end{aligned}
\end{equation}
where $S_t^{n_m}$ represents the $t$-th session data of neighbour $n_m, t \in  [1, T ]$, and $\mathcal{N}_{{\bar{u}}}^{T+1}$ is the set of neighbours of target user ${\bar{u}}$. Here, we denote $h_{{\bar{u}},k}'$ as the encoding result of the $k$-th mask sequence of target user ${\bar{u}}$ in Session $T$+1 into $H_{\bar{u}}$, $\intercal$ as the transpose operation, and $|I|$ as the total number of items.

\subsection{Model Training}
We treat the SSR task as a classification task that allows the model to learn which items the user is interested in. We use the maximum likelihood estimation to train the model and  gradient descent to optimize the model:
\begin{equation}
    \begin{aligned}
        -\sum_{u\in U} \sum_{t=2}^{T} \sum_{k=1}^{n-1} log&(p( Target_k| MaskedInput_k;\\&\{S_{t<T}^{n_m},n_m\in \mathcal{N}_{{\bar{u}}}^{T+1}\}))
    \end{aligned}
\end{equation}

\section{Experiments}
In this section, we describe the experimental setup including the used real-world datasets and the comparative baseline methods as well as the used evaluation metrics. We also introduce the implementation details to ensure the reproduction of our model. Finally, we provide  the quantitative analysis of the results, the ablation study and the hyper-parameter analysis.

\subsection{Experimental Setup}
\subsubsection{Datasets} We conducted experiments on the following four commonly used real-world datasets, for which descriptive statistical information is shown in Table~\ref{tab:dataset}. 





\textit{Douban music and Douban movie}\footnote{\label{foot01}http://www.douban.com}. Douban is a popular communication platform including the inforamtion of music and movie. Users can rate and communicate about music and movie. The dataset includes user's ratings of music and movie and the corresponding timestamps, as well as social network information.

\textit{Extended epinions}\footnote{http://www.trustlet.org/epinions.html}. This dataset contains information about users' trust and distrust. Users usually rate other users based on the quality of their reviews of the item.

\textit{Yelp}\footnote{https://www.yelp.com/dataset}. A very popular online review platform where users can review local businesses (e.g. restaurants and shops), providing content-rich social networking information. 

Following~\cite{song2019session}, we partitioned a dataset by segmenting user behavior into weekly sessions as the train, validation, and test sets. Specifically, sessions within the last $s$ weeks were retained for evaluation. For the Douban music, Douban movie, Extended Epinions, and Yelp datasets, we selected $s$ values of 26, 26, 14, and 24, respectively. Additionally, we filtered out items that did not appear in the training set to ensure consistency across sets. We randomly partitioned the sessions utilized for the testing and evaluation, and evenly distributed them.

\begin{table}[ht]
    \renewcommand\arraystretch{1.3}
    \caption{Statistical information on the Datasets}\label{tab:dataset}
    \centering
    \resizebox{\linewidth}{!}{
        \fontsize{9}{10}\selectfont
        \begin{tabular}{m{25mm}<{\centering} *{4}{m{12mm}<{\centering}}}
            \toprule[1.3pt]
            & Douban music & Douban movie & Extended epinions & Yelp \\
           \Xhline{0.25pt}
            \# Users & 9,501 & 26,511 & 7,626 & 95,591 \\
            \# Items & 3,945 & 12,591 & 5,959 & 12,077 \\
            \# Events & 258,398 & 2,777,070 & 123,344 & 674,521 \\
            \# Social links & 18,487 & 77,197 & 170,981 & 1,454,496 \\
            Start Date & 11/01/2008 & 11/01/2008 & 27/09/2000 & 01/01/2009 \\
            End Date & 21/01/2017 & 21/01/2017 & 05/07/2004 & 10/15/2016 \\
            Avg. friends/user & 1.95 & 2.91 & 22.42 & 15.22 \\
            Avg. events/user & 27.20 & 104.75 & 16.17 & 7.06 \\
            Avg. session length & 3.82 & 3.86 & 3.22 & 2.88 \\
            \bottomrule[1.3pt]
        \end{tabular}
    }\vspace*{-1ex}
\end{table}

\begin{table*}
    \centering
    \renewcommand\arraystretch{1.5}
    \caption{Performance comparison of TEGAARec versus other baseline models in terms of two metrics (\%)}
    \small
    \setlength{\tabcolsep}{6pt} 
    \begin{tabular}
     {m{20mm}<{\centering} p{10mm}<{\centering} p{10mm}<{\centering} p{10mm}<{\centering} p{10mm}<{\centering} p{10mm}<{\centering} p{10mm}<{\centering} p{10mm}<{\centering} p{10mm}<{\centering}}
        \toprule
        \multirow{2}{*}{{\makecell{\textbf{Model}}}} & \multicolumn{2}{c}{{\makecell{\textbf{Douban music}}}} & \multicolumn{2}{c}{{\makecell{\textbf{Douban movie}}}} & \multicolumn{2}{c}{{\makecell{\textbf{Extended epinions}}}} & \multicolumn{2}{c}{{\makecell{\textbf{Yelp}}}}\\
        \cline{2-9}        
        & \textbf{R@20} & \textbf{N@20} & \textbf{R@20} & \textbf{N@20} & \textbf{R@20} & \textbf{N@20} & \textbf{R@20} & \textbf{N@20} \\
        \hline
        NextItNet & 15.26 & 7.360 & 14.70 & 8.120 & 14.82 & 9.240 & 7.590 & 3.030\\
        
        NARM & 16.51 & 7.980 & 14.59 & 7.870 & 17.27 & 10.36 & 7.690 & 3.050\\

        STAMP & 15.18 & 7.640 & 14.25 & 8.060 & 15.24 & 9.560 & 7.340 & 3.100\\
        
        SR-GNN & 16.85 & 8.280 & 14.82 & 8.350 & 16.71 & 10.13 & 7.690 & 3.090\\
        
        GCE-GNN & 35.68 & 22.52 & 19.91 & 9.647 & 15.41 & 10.56 & 30.99 & 13.78\\
        
        HIDE & 27.08 & 18.16 & 20.15 & 9.240 & 21.08 & 10.63 & \uline{33.97} & 13.78\\
        
        Attn-mixer & 25.56 & 17.03 & 20.11 & 11.15 & \uline{23.50} & 11.77 & 22.79 & 13.50\\

        \hline
        
        SERec & 19.21 & 12.76 & 18.50 & 9.300 & 18.31 & 12.91 & 15.89 & 12.78\\
        
        DGRec & \uline{37.77} & 29.47 & 18.72 & 19.50 & 13.13 & 14.88 & 18.84 & 19.54\\

        GESU & 24.15 & 22.02 & 14.97 & 12.46 & 14.25 & 15.93 & 18.39 & 19.20\\
        
        GNNRec & 21.07 & \uline{31.14} & \uline{27.75} & \uline{22.66} & 21.21 & \uline{20.41} & 11.28 & \uline{21.32}\\



        \hline
        
        \textbf{TEGAARec} & \textbf{81.28} & \textbf{47.50} & \textbf{59.84} & \textbf{34.78} & \textbf{47.46} & \textbf{33.95} & \textbf{73.12} & \textbf{51.77}\\
        \bottomrule
    \end{tabular}
    \label{tab:t2}
\end{table*}

\subsubsection{Comparative Methods}
We selected a number of representative works for comparison including:
\begin{itemize}
    \item \textbf{NextItNet}\cite{yuan2019simple} -- A CNN-based model with long-range modeling capabilities. It expands the convolutional layers to increase the receptive field and introduces residual networks into multiple convolutional layers.
    \item \textbf{NARM}\cite{li2017neural} -- A RNN-based model. It better captures users' sequential behaviours and personal intentions by introducing attention mechanisms into GRUs.
    \item \textbf{STAMP}\cite{liu2018stamp} -- An attention-based model. It captures both users' long-term and short-term interests using the attention mechanism.
    \item \textbf{SR-GNN}\cite{wu2019session} -- A GNN-based model. It converts sequential problems into graph problems by constructing sequences of conversations as directed graphs and then learns the complex transitions between items through gated neural networks.
    \item \textbf{GCE-GNN}\cite{wang2020global} -- A GNN-based model. It learns information about item-transitions from other sessions through the construction of a global graph, which are further fed into the graph encoder.
    \item \textbf{HIDE}\cite{li2022enhancing} -- This method delineates potential shifts in interest from diverse perspectives by generating a hypergraph for each session. Following this, both micro and macro methodologies are employed to discern the underlying purpose of each clicked item.
    \item \textbf{Attn-Mixer}\cite{zhang2023efficiently} -- An attention-based model. It proposes a multi-level attention hybrid network and uses multi-level user intentions to achieve multi-level reasoning for item transitions.
    \item \textbf{SERec}\cite{chen2021efficient} -- A social recommendation model. It learns the user-item representations using heterogeneous graph neural networks to capture item transitions across sessions.
    \item \textbf{DGRec}\cite{song2019session} -- A social recommendation model. It uses RNNs and graph attention networks to model users' dynamic interests and the influence of their friends.
    \item \textbf{GESU}\cite{GESU} -- A social recommendation model. It captures the rapid user interest changes using GGNN and multi-head attention mechanisms, and aggregates social influence from friends using graph attention mechanisms.
    \item \textbf{GNNRec}\cite{liu2023gnnrec} -- A social recommendation model. It improves performance of DGRec by constructing session graphs and uses GRU to model complex transitions between items.
\end{itemize}
\subsubsection{Evaluation Metrics}
We used two popular ranking-based metrics for our evaluation\cite{song2019session}: \textbf{R@K}(Recall@K) and \textbf{N@K}(NDCG@K). For R@K, we use K=\{10, 20\}, while for N@K, we set K=20 following \cite{song2019session}.  

\subsubsection{Implementation Details}
We use the Pytorch framework \cite{paszke2019pytorch} to build our models, and all experiments are run on two Tesla V100 GPUs. For faster model training convergence,
we used the warm-up technique~\cite{goyal2017accurate} to help the model parameters converge faster for learning rate and training epochs. Specifically, we use the grid search method to search for the following hyper-parameters: learning rate in \{0.01, 0.005, 0.0001, 0.00005\}, number of neighbours in \{25, 50\}, number of LMP in \{5, 15, 25\}, number of layers of TEGAA modules in \{1, 3, 5\}, number of warm-up learning steps in \{5, 10, 20\}, and the number of early stop tolerances in \{10, 20\}. We used a uniform batch size of 50 and the Adam~\cite{kingma2014adam} optimiser. For all baseline model experiments, we used their default hyper-parameters, and we chose the highest value between our experimental results and those reported in the original paper for other baseline models.

\subsection{Quantitative Results}
Table~\ref{tab:t2} presents a performance comparison between our proposed TEGAARec model and selected baseline models. The experimental findings reveal that, apart from the Attn-mixer model, models incorporating social influence such as DGRec and GNNRec outperform those lacking social influence. This underscores the beneficial impact of social influence on recommendation performance, particularly evident in datasets like Yelp with numerous users and dense social networks. Moreover, GCE-GNN model gains a significant advantage by leveraging global graph construction using other session information, further validating the benefits of incorporating such data in recommendation tasks.
In contrast, the Attn-mixer model, which enhances inference through multi-level user intent modeling, achieves competitive results, highlighting the advantages of employing attention mechanisms.
Our proposed TEGAARec model effectively identifies users with genuinely similar preferences to the target user by leveraging Like-minded Peers (LMP) mining in the recommendation data. Additionally, it accurately models the dynamic interests of users. The results demonstrate a substantial performance improvement compared to other state-of-the-art works.

\begin{table}[h]
    \centering
    \renewcommand\arraystretch{1.5} 
    \caption{Performance comparison of TEGAARec variants with the original model in terms of two metrics (\%)}
    \setlength{\tabcolsep}{4pt} 
    \normalsize
    \resizebox{0.49\textwidth}{!}{
        \begin{tabular}{m{35mm}<{\centering}@{\hspace{3mm}}m{20mm}<{\centering}@{\hspace{3mm}}m{14mm}<{\centering}@{\hspace{3mm}}m{14mm}<{\centering}@{\hspace{3mm}}m{14mm}<{\centering}@{}}
            \toprule
            \multirow{2}{*}{{\makecell{\textbf{Model}}}} & \multicolumn{2}{c}{{\textbf{Douban music}}} & \multicolumn{2}{c}{{\textbf{Yelp}}}\\
            \cline{2-5}  
            & \textbf{R@20} & \textbf{N@20} & \textbf{R@20} & \textbf{N@20}\\
            \hline
            TEGAARec w/o LMP & 22.29 & 19.23 & 12.38 & 16.28\\
            TEGAARec w/o GAL & 66.42 & 39.57 & 67.08 & 45.94\\
            TEGAARec w/o SF & 79.57 & 40.94 & 70.03 & 48.62\\
            TEGAARec w/ PE & 80.03 & 44.29 & 70.64 & 49.68\\
            TEGAARec w/o ULI & 78.95 & 40.27 & 71.55 & 50.19\\
            TEGAARec w/o ALI & 15.15 & 17.65 & 5.49 & 13.05\\
            \hline
            \textbf{TEGAARec} & \textbf{81.28} & \textbf{47.50} & \textbf{73.12} & \textbf{51.77}\\
            \bottomrule
        \end{tabular}
    }
    \label{tab:t3}
\end{table}

\subsection{Ablation Studies}

To assess the effectiveness of each key component in the TEGAARec model, we conducted ablation experiments from six different perspectives. These experiments were carried out on the Douban music and Yelp datasets, and the results are summarized in Table~\ref{tab:t3}:
\begin{itemize}
\item \textbf{TEGAARec w/o LMP}: This variant utilizes only the social relationships provided by the dataset and excludes the use of Like-minded Peers (LMP).
\item \textbf{TEGAARec w/o GAL}: It directly concatenates dynamic interest influences of different neighbours without employing graph attention aggregation.
\item \textbf{TEGAARec w/o SF}: It excludes the social friends of the target user while keeps the LMP only as the neighbours.
\item \textbf{TEGAARec w/ PE}: This variant incorporates position encoding into the TEGAA module.
\item \textbf{TEGAARec w/o ULI}: It excludes the target user’s long-term interest embedding from the model.
\item \textbf{TEGAARec w/o ALI}: This variant removes all user long-term interest embeddings from the model.
\end{itemize}

The experimental results demonstrate that the presence of Like-minded Peers (LMP), the Graph Attention Aggregation Layer, and the inclusion of long-term interests of target users and their neighbours are pivotal factors influencing the model's performance. Removal or replacement of any of these components, as observed in TEGAARec w/o LMP, TEGAARec w/o GAL, and TEGAARec w/o ALI, respectively, leads to a significant degradation in performance.
Interestingly, the absence of significant performance improvement in TEGAARec w/ PE suggests that users' short-term interests are more accurately captured through a collection of interaction items over a brief period, without the need for sequence and position information, which may be redundant.
Also, the removal of social friends has a small influence on the model's performance. This proves the contribution of our designed LMP to some extent especially in the case of sparse social friends.
Additionally, TEGAARec w/o ULI resulted in a slight performance degradation. This observation could be attributed to the fact that the target user's long-term interests are inherently incorporated when they act as a neighbour to other users.

\subsection{Hyper-parameter Sensitivity Analysis}
During the model training phase, we employed a grid search technique to optimize the hyperparameters. However, considering the large number of hyperparameters involved, we conducted the search on a reduced subset to maintain training efficiency. To further investigate the influence of hyperparameters on model performance, we selected specific key hyperparameters for sensitivity analysis. Consistent with the ablation experiments, we also performed hyper-parameter sensitivity experiments on the Douban music and Yelp datasets.

\subsubsection{Impact of Embedding Dimension}
\begin{figure}[ht]
	\centering
	\includegraphics[width=1.0\linewidth,keepaspectratio]{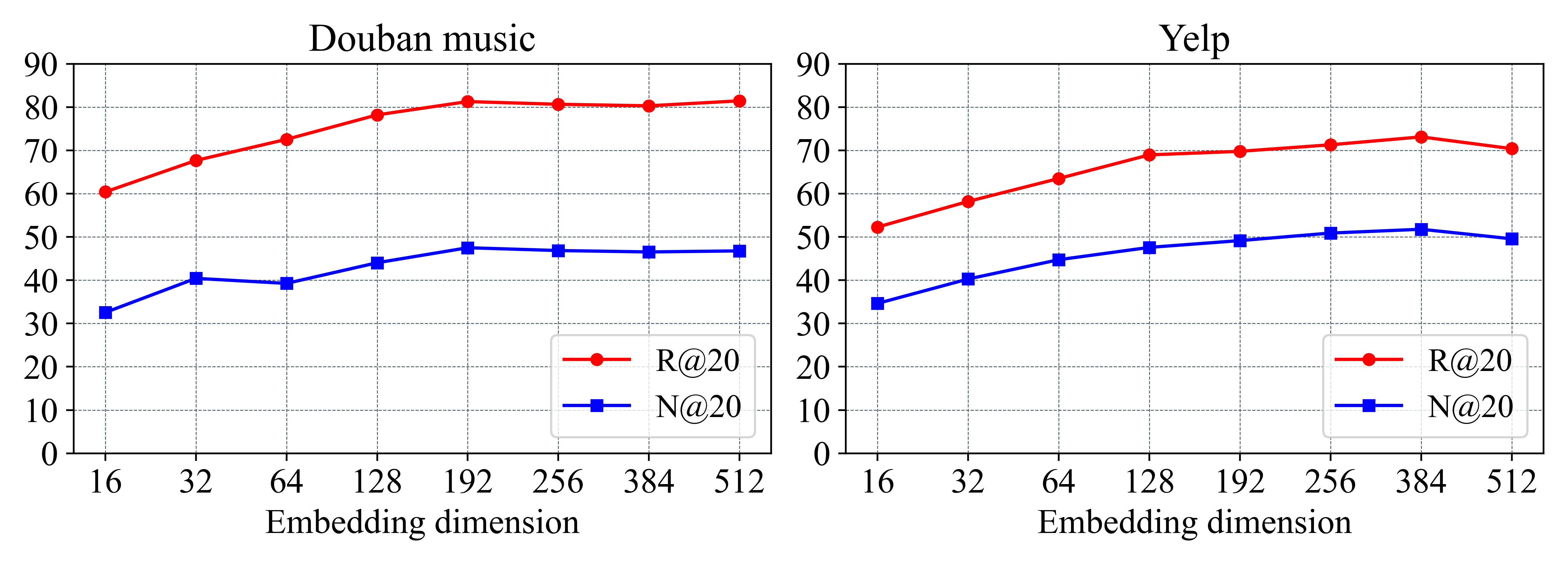}
	\caption{Effect of different embedding dimension sizes on model performance}
    \label{fig:5}
\end{figure}

To assess the effect of different embedding dimensions (utilized for both user and item embeddings) on model performance, we experimented with embedding dimensions ranging from 16 to 512. The results, depicted in Fig. \ref{fig:5}, reveal that the model's performance improvement becomes limited when the embedding dimension exceeds 128 on both datasets, with a marginal decreasing effect thereafter.

\subsubsection{Impact of Transformer Encoder Layer}
\begin{figure}[ht]
	\centering
	\includegraphics[width=1.0\linewidth,keepaspectratio]{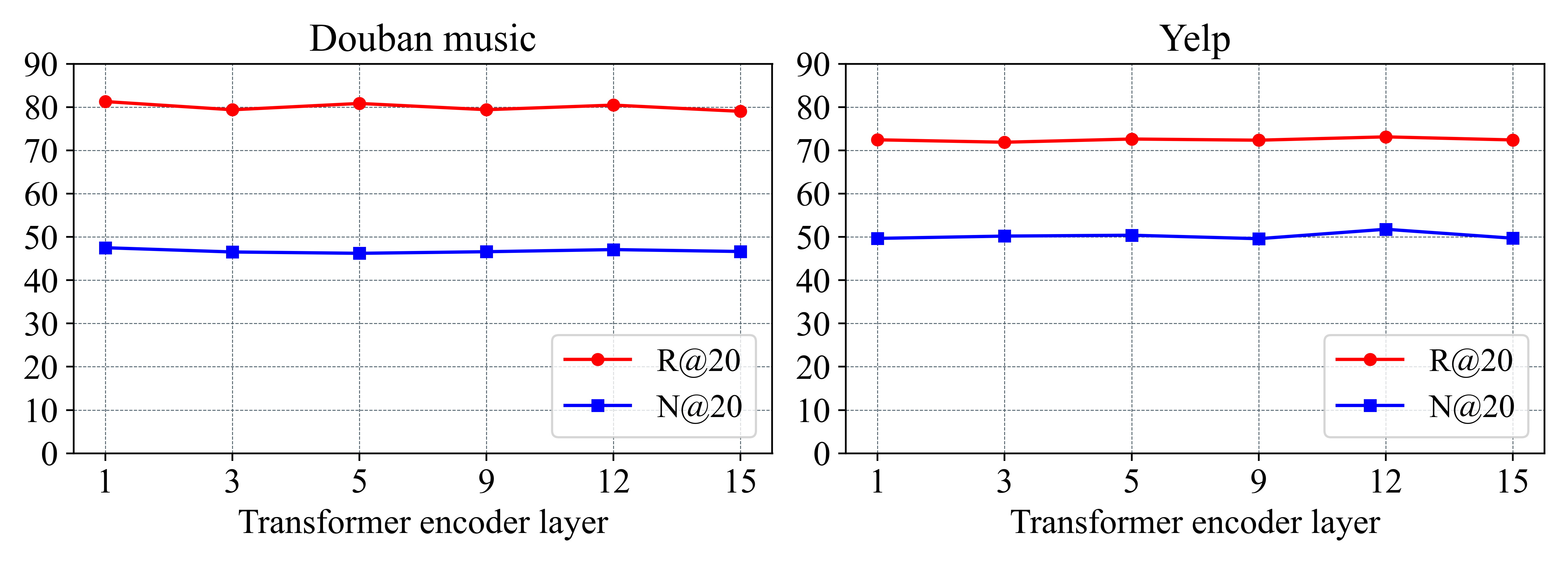}
	\caption{Effect of different number of Transformer encoder layers on model performance}
    \label{fig:6}
\end{figure}

To evaluate the impact of varying Transformer Encoder layers on model performance, we conducted experiments with different layer numbers ranging from 1 to 15. The results, as illustrated in Fig. \ref{fig:6}, indicate that employing different Transformer Encoder layers does not substantially influence the model's performance on both datasets. Surprisingly, our findings suggest that a single layer of Transformer Encoder is adequate for achieving effective modeling.

\subsubsection{Impact of the number of LMP users}
\begin{figure}[ht]
	\centering
	\includegraphics[width=1.0\linewidth,keepaspectratio]{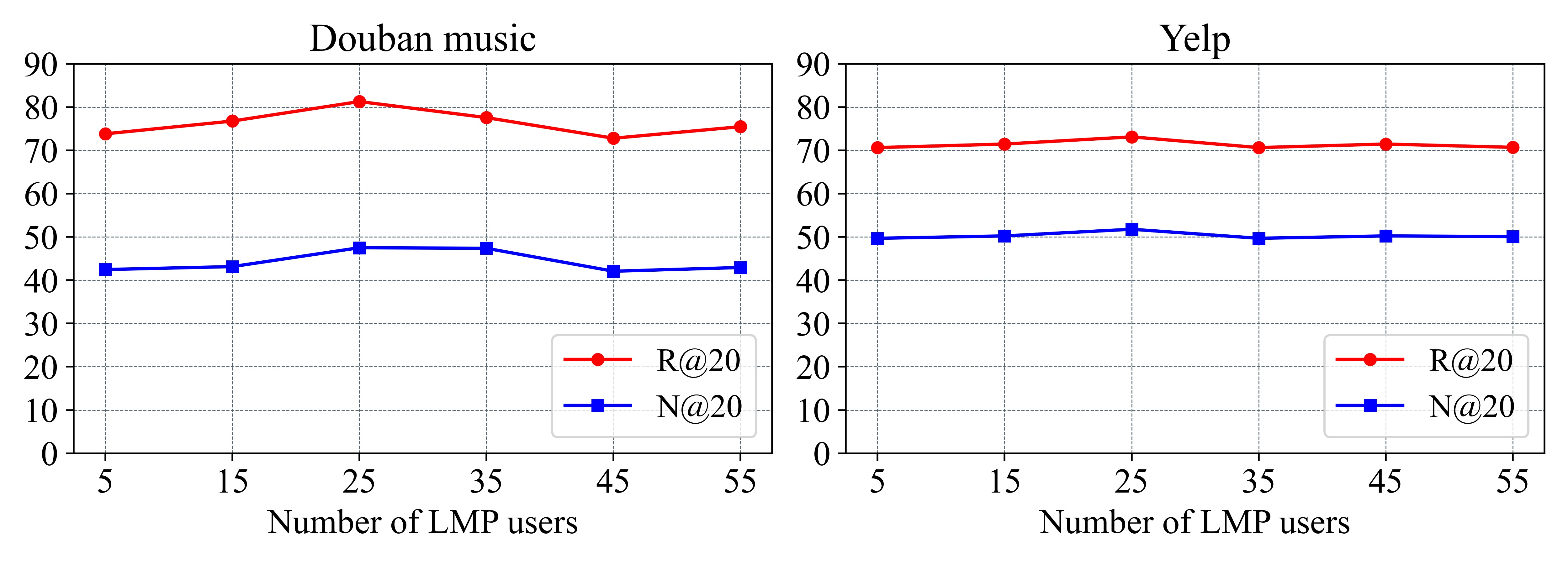}
	\caption{Effect of different number of LMP users on model performance}
    \label{fig:7}
\end{figure}

To evaluate the impact of different LMP users numbers on model performance, we conducted experiments with the numbers ranging from 5 to 55. As depicted in Fig. \ref{fig:7}, the results indicate that increasing the number of LMP users enhances the model performance when the number of samples is less than 25 on both datasets. However, when the number of LMP samples exceeds 25, a slight degradation in model performance is observed, possibly attributable to the introduction of noise from an excessive number of LMP users.

\subsubsection{Impact of the number of social friend samples}
\begin{figure}[ht]
	\centering
	\includegraphics[width=1.0\linewidth,keepaspectratio]{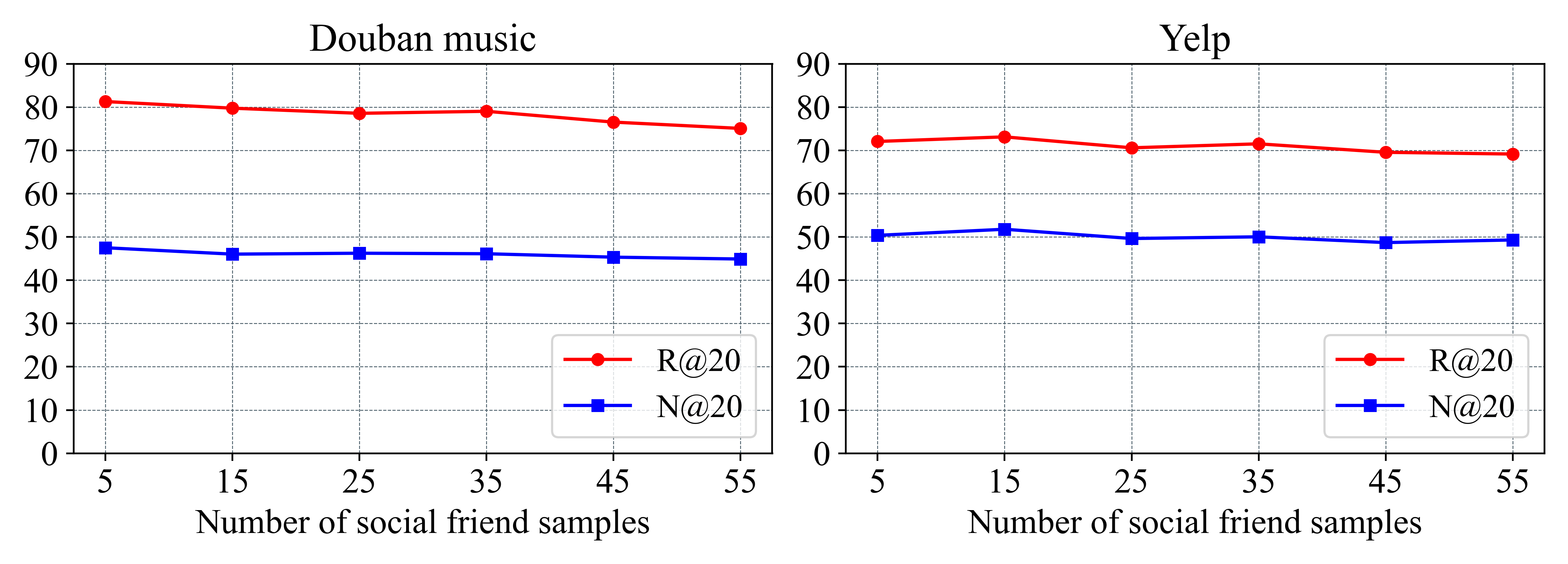}
	\caption{Effect of different number of social friend samples on model performance}
    \label{fig:8}
\end{figure}

To evaluate the impact of varying social friend sampling numbers on model performance, we experimented with numbers ranging from 5 to 55. As illustrated in Fig.\ref{fig:8}, the results indicate that employing different numbers of social friend samples does not significantly affect the model performance on both datasets. This observation suggests that the influence from social friends is limited especially when it occurs with the problems of social friends sparsity. This finding is further corroborated by our ablation experiments.

\begin{figure*}[h]
	\centering
\includegraphics[width=0.95\textwidth]{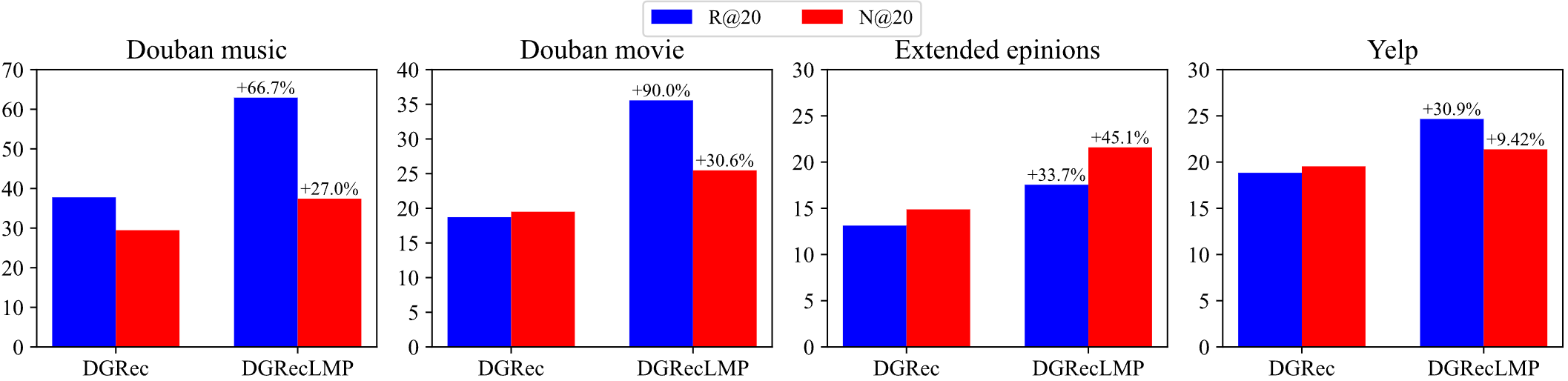}
	\caption{Performance comparison of the DGRec model before and after the introduction of LMP.}
	\label{fig:3}
\end{figure*}

\begin{figure*}[h]
	\centering
	\includegraphics[width=0.95\textwidth]{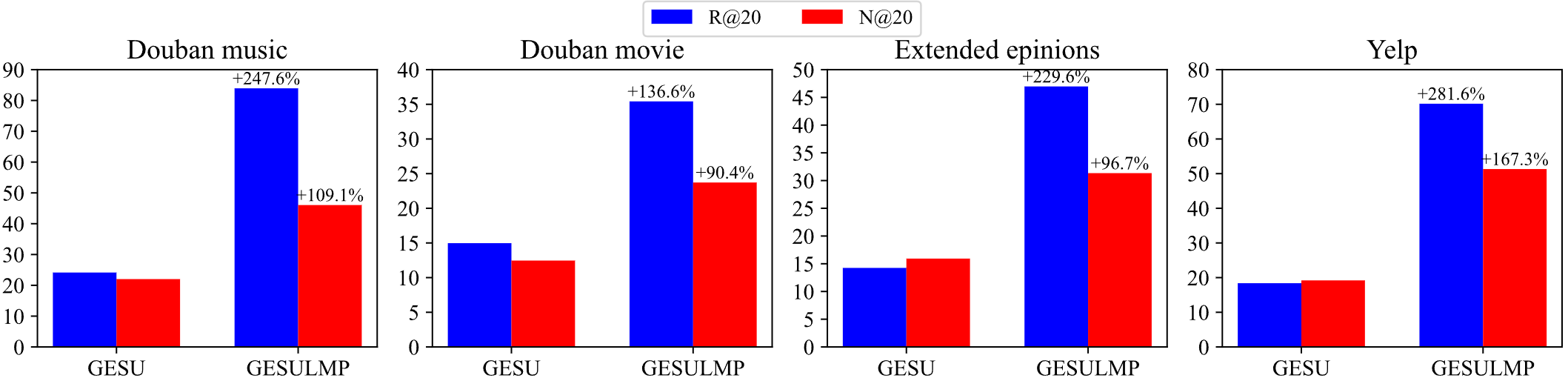}
	\caption{Performance comparison of the GESU model before and after the introduction of LMP.}
	\label{fig:4}
\end{figure*}

\begin{table*}
    \centering
    \renewcommand\arraystretch{1.5}
    \caption{Performance comparison of TEGAARec versus two other baseline models incorporating LMP in terms of two metrics (\%)}
    \small
    \setlength{\tabcolsep}{6pt} 
    \begin{tabular}
     {m{20mm}<{\centering} p{10mm}<{\centering} p{10mm}<{\centering} p{10mm}<{\centering} p{10mm}<{\centering} p{10mm}<{\centering} p{10mm}<{\centering} p{10mm}<{\centering} p{10mm}<{\centering}}
        \toprule
        \multirow{2}{*}{{\makecell{\textbf{Model}}}} & \multicolumn{2}{c}{{\makecell{\textbf{Douban music}}}} & \multicolumn{2}{c}{{\makecell{\textbf{Douban movie}}}} & \multicolumn{2}{c}{{\makecell{\textbf{Extended epinions}}}} & \multicolumn{2}{c}{{\makecell{\textbf{Yelp}}}}\\
        \cline{2-9}        
        & \textbf{R@20} & \textbf{N@20} & \textbf{R@20} & \textbf{N@20} & \textbf{R@20} & \textbf{N@20} & \textbf{R@20} & \textbf{N@20} \\
        \hline

        DGRec & 37.77 & 29.47 & 18.72 & 19.50 & 13.13 & 14.88 & 18.84 & 19.54\\

        GESU & 24.15 & 22.02 & 14.97 & 12.46 & 14.25 & 15.93 & 18.39 & 19.20\\


        DGRecLMP & 62.93 & 37.43 & 35.57 & 25.47 & 17.55 & 21.59 & 24.66 & 21.38\\

        GESULMP & \textbf{83.94} & 46.05 & 35.42 & 23.72 & 46.97 & 31.34 & 70.17 & 51.32\\

        \hline
        \textbf{TEGAARec} & 81.28 & \textbf{47.50} & \textbf{59.84} & \textbf{34.78} & \textbf{47.46} & \textbf{33.95} & \textbf{73.12} & \textbf{51.77}\\
        \bottomrule
    \end{tabular}
    \label{tab:t3}
\end{table*}

\subsection{Scalability of LMP}
To further explore the influence of LMP, we conducted the extended experiments on the DGRec and GESU models to verify whether LMP can work well on other models on the four datasets. We denote the improved DGRec and GESU models by introducing LMP as DGRecLMP and GESULMP, respectively.
Fig.~\ref{fig:3} shows the performance comparison of the DGRec model before and after the introduction of LMP. The experimental results show that 
the introduction of LMP helps to improve the performance on all four datasets, especially on the Douban music and Douban movie with a 66.7\% and 90\% improvement for Recall@20, respectively. 
Similarly, Fig.~\ref{fig:4} shows the performance comparison of the GESU model before and after the introduction of LMP. The experimental results show that GESULMP achieves more significant improvements compared with the DGRec model. Moreover, on the metrics of the Douban music dataset, the performance of the GESULMP model achieves quite excellent results, with a 247.6\% and 109.1\% improvement for the metric Recall@20 and NDCG@20, respectively. By analyzing the model structure, both GESU and TEGAARec use the multi-head attention module, which has more powerful data modeling capabilities, thus achieving the superior results compared to the GRU network used in the DGRec model. These experimental results further verify the scalability of the LMP concept.

In addition, we compare the results of the improved DGRec model and GESU model, i.e.,  DGRecLMP and GESULMP, with our TEGAARec model on four dataset in Table \ref{tab:t3}. 
Although the GESULMP model achieves very competitive performance by introducing LMP, our proposed TEGAARec model still maintains a large advantage, especially in modeling large-scale data more efficiently, like the Douban movie dataset, which is an order of magnitude larger than the other three datasets.

\section{Conclusions}

In this paper, we introduced the concept of ``Like-minded Peers'' (LMP) to tackle the problem of social friend sparsity and promote the identification of users who genuinely share similar preferences with the target users. Additionally, we proposed the Transformer Encoder with Graph Attention Aggregator Recommendation (TEGAARec) model, which offers improved modeling of dynamic interests and social influences among users. Both the long-term and short-term interests for target users and LMP users are captured and merged to get the final representations of target users.
The substantial performance enhancement achieved by our model across four real-world datasets from diverse domains underscores the effectiveness and superiority of our approach.

For future work, we aim to delve deeper into two key aspects. Firstly, we plan to explore more fine-grained mining of LMP, including considerations of the relative importance of different LMP to a target user. Secondly, we intend to investigate methods for better modeling the impact of different items within a session, thereby enhancing the overall recommendation accuracy and user experience.

\section{Acknowledgement}
The authors wish to thank the Projects of Natural Science Foundation of Inner Mongolia under Grant No.2024MS06014, Inner Mongolia Science \& Technology Plan under Grant No.2021GG0164, Natural Science Foundation of China under Grant No.61962039, 62162046, 62362054, 62366038, Inner Mongolia Engineering Lab of Cloud Computing and Service Software and Inner Mongolia Engineering Lab of Big Data Analysis Technology,the Engineering Research Center of Ecological Big Data, Ministry of Education, China.

\bibliographystyle{cas-model2-names}

\bibliography{cas-refs}



\end{document}